\documentclass{article}
\usepackage{spconf,amsmath,graphicx}
\usepackage{amsmath}
\usepackage{amssymb}
\usepackage{mathtools}
\usepackage{tikz}
\usepackage{hyperref}

\newcommand{\M}{\mathcal{M}}
\newcommand{\R}{\mathbb{R}}
\newcommand{\N}{\mathcal{N}}
\newcommand{\E}{\mathbb{E}}
\newcommand{\Ee}{\E_{\text{MSE}}}
\newcommand{\Eh}{\hat{\E}_{\text{MSE}}}
\newcommand{\zh}{\hat{z}_\M}

\title{Optimal measurement budget allocation for particle filtering}
\name{Antoine ASPEEL, Amaury GOUVERNEUR, Rapha\"el M. JUNGERS and Beno\^it MACQ\thanks{A.A and A.G contributed equally. A.A. is supported by the Walloon Region, its grant is RW-DGO6-Biowin-Bidmed. R.J. is a FNRS Research Associate. He is supported by the Walloon Region and the Innoviris Foundation. \textcopyright\ 2020 IEEE. Personal use of this material is permitted. Permission from IEEE must be obtained for all other uses, in any current or future media, including reprinting/republishing this material for advertising or promotional purposes, creating new collective works, for resale or redistribution to servers or lists, or reuse of any copyrighted component of this work in other works.}}
\address{ICTEAM Institute, UCLouvain, Avenue Georges Lemaître 4-6, Louvain-la-Neuve, Belgium}
\begin{document}
%
\maketitle
%


\begin{abstract}
Particle filtering is a powerful tool for target tracking. When the budget for observations is restricted, it is necessary to reduce the measurements to a limited amount of samples carefully selected. A discrete stochastic nonlinear dynamical system is studied over a finite time horizon. The problem of selecting the optimal measurement times for particle filtering is formalized as a combinatorial optimization problem. We propose an approximated solution based on the nesting of a genetic algorithm, a Monte Carlo algorithm and a particle filter. Firstly, an example demonstrates that the genetic algorithm outperforms a random trial optimization. Then, the interest of non-regular measurements versus measurements performed at regular time intervals is illustrated and the efficiency of our proposed solution is quantified: better filtering performances are obtained in $87.5\%$ of the cases and on average, the relative improvement is $27.7\%$.
\end{abstract}
\section{Introduction}
Stochastic nonlinear dynamical systems have shown their ability to model a number of real-world problems \cite{arnold2012stochastic}. Particle filtering is an efficient approach to estimate the state of such systems from a set of noisy measurements \cite{arulampalam2002tutorial}. This tool has been largely used, among others, in computer vision \cite{hue2001particle,cho2006real,andreasson2005localization}. In practice, performing measurements may be difficult due to energy consumption, economical constraints or health hazards. For instance, in tumor tracking based on X-ray images, the number of X-ray acquisitions should be minimized in order to limit patients' exposure to harmful radiations \cite{sharp2004prediction}.


Under such constraints, the problem is then to select the best moments to measure the system \textit{a priori}, i.e. before any measurement acquisition. In other words, one has a measurement budget and has to choose when to acquire measurements. The optimality criterion is to minimize the expected filtering \textit{mean squared error} (MSE) over the complete time horizon. 

In the particular case of linear systems subject to Gaussian noise processes, the selection of optimal measurement times over a finite time horizon has been studied using the Kalman filtering framework, in both discrete \cite{aspeel2019optimal} and continuous-time \cite{sano1970measurement,aksenov2019optimal} settings. However, more general formulations have received little attention in the literature. This paper addresses the problem of providing optimal measurement times in the discrete-time nonlinear case with perturbation and measurement noise processes following arbitrary distributions. Our approach relies on particle filtering, the efficiency of which has been widely demonstrated in nonlinear dynamical systems \cite{thrun2002particle,gustafsson2002particle,rahni2011particle}.

The two main contributions of this paper are (i) to propose an efficient algorithm to solve the problem of optimal measurement times selection and (ii) to show the interest of non-regular measurements in particle filtering.

This paper is organised as follows, section \ref{sec:materianlsAndMethods} presents how to implement particle filtering with intermittent measurements (subsection \ref{sec:IPF}); the criterion to select a good set of measurement times (subsection \ref{sec:optimalIPF}) and how to compute them (subsections \ref{sec:MC} and \ref{sec:ga}). An example is presented and discussed in section \ref{sec:resultsAndDiscussions}. Finally, section \ref{sec:conclusion} concludes and discusses possible improvements and perspectives.

A \textsc{Matlab} (MathWorks, Natick, Massachusetts, USA) implementation of all the presented algorithms and the code that generate all figures are available on \textsc{GitHub} at\\
\href{https://github.com/AmauryGouverneur/Optimal_Measurement_Budget_Allocation_For_Particle_Filtering}{\textsc{github.com/AmauryGouverneur/Optimal\_\\ Measurement\_Budget\_Allocation\_For\_Particle\_\\ Filtering}}.

\section{Materials and methods}\label{sec:materianlsAndMethods}

\subsection{Intermittent particle filter}\label{sec:IPF}
A discrete stochastic nonlinear dynamical system describes the evolution of a state $x(t)$ over the finite time horizon $t=0,\dots,T$. One wants to estimate a quantity $z(t)$ related to $x(t)$ and has access to previously acquired noisy measurements $y(t)$ of $x(t)$. Measurements are not available at each time step. More formally, a measurement $y(t)$ is only available for $t\in\M$, where $\M\subseteq \{0,\dots,T\}$ is of size $N$, i.e. $|\M|=N$. This is modelled as
\begin{alignat}{3}
x(t+1) &=&&\ f_t(x(t),w(t))  &\text{\ \ for\ \ } & t=0,\dots,T-1, \label{eq:model:x}\\
y(t)   &=&&\ g_t(x(t),v(t))  &\text{\ \ for\ \ } & t\in\M, \label{eq:model:y}\\
z(t)   &=&&\ h_t(x(t)) &\text{\ \ for\ \ } & t=0,\dots,T,\label{eq:model:z}\\ 
x(0)   &\sim&&\ \mathcal{F},\label{eq:model:x0}
\end{alignat}
where $x(t)\in\R^n$, $y(t)\in\R^m$ and $z(t)\in\R^p$. In addition, $w(t)$ and $v(t)$ are random processes with known probability density functions. Functions $f_t(\cdot,\cdot)$, $g_t(\cdot,\cdot)$ and $h_t(\cdot)$ are known and have compatible dimensions. The initial state $x(0)$ follows a known distribution $\mathcal{F}$.

For instance, in a tumor tracking problem based on X-ray images, $x(t)\in\R^6$ can be a state vector containing the tumor’s position and velocity in the 3-dimensional space, $y(t)\in\R^2$ can be the 2-dimensional projection of the target and $z(t)\in\R^3$ the position of the mass center in the 3-dimensional space.

An estimate $\zh(t)=PF[\{y(\tau)\}_{\tau\leq t,\ \tau\in\M}]$ of $z(t)$ based on previous measurements $\{y(\tau)\}_{\tau\leq t,\ \tau\in\M}$, can be computed by a particle filter $PF[\cdot]$. It is the expectation of the estimated probability density function.

Essentially, a particle filter algorithm alternates between (i) a \textit{prediction step} (also called mutation), used to estimate the state at the next time step from the estimate at the current step; and (ii) a \textit{correction step} (also called selection) that updates the state estimation to incorporate the information acquired in the last measurement. To deal with intermittent measurements, the correction step (ii) is skipped when no measurement is available, i.e. when $t\notin \M$.

In this paper, we use the sampling importance resampling particle filter (see Algorithm 4 in \cite{arulampalam2002tutorial}).


\subsection{Optimal intermittent particle filter}\label{sec:optimalIPF}
The optimal intermittent particle filter is the particle filter for which the set of $N$ measurement times $\M$ minimizes the expected filtering MSE. This is formalized as
\begin{align}\label{eq:problem}
\min_{\M\subseteq\{0,\dots,T\}} \Ee[\M] \coloneqq \E\left[ \dfrac{1}{T+1}\sum_{t=0}^T  \| z(t)-\zh(t) \|^2 \right]\nonumber \\
\text{subject to}\ |\M|=N \text{\ and\ equations\ (\ref{eq:model:x})\ to\  (\ref{eq:model:x0})},
\end{align}
where $\zh(t)$ is obtained from the particle filter and $\|\cdot\|$ is the Euclidean norm (note that it could be any other norm). The expectation is on the random variables $x(t)$ and $v(t)$. The dependency of the cost function $\Ee[\M]$ on these two quantities can be expressed explicitly using equations (\ref{eq:model:y})-(\ref{eq:model:z}), it gives
\begin{small}
\begin{align}
\E\left[\frac{1}{T+1}\sum_{t=0}^T\| h_t(x(t)) - PF[\{g_\tau(x(\tau),v(\tau))\}_{\tau\in\M,\tau\leq t}] \|^2\right].\nonumber
\end{align}
\end{small}
Solving problem (\ref{eq:problem}) yields the best measurement times \textit{a priori}, i.e. before any measurement acquisition.

\subsection{Monte Carlo algorithm}\label{sec:MC}
A first challenge to solve problem (\ref{eq:problem}) is to compute the expectation. To tackle this problem, we estimate this expectation using a Monte Carlo approach. For a given set of measurement times $\M$, one can simulate $K$ realisations $\{x^k(t)\}^{k=1,\dots,K}_{t=0,\dots,T},\{y^k(t)\}^{k=1,\dots,K}_{t\in\M}$ and $\{z^k(t)\}^{k=1,\dots,K}_{t=0,\dots,T}$ drawn according to (\ref{eq:model:x})-(\ref{eq:model:x0}). From each simulated sequence of measurements $\{y^k(t)\}_{t\in\M}$, the particle filter computes estimates $\{\zh^k(t)\}_{t=0,\dots,T}\coloneqq \{PF[\{y^k(\tau)\}_{\tau\leq t,\tau\in\M}]\}_{t=0,\dots,T}$ of $\{z^k(t)\}_{t=0,\dots,T}$. These quantities are used to estimate the expectation in problem (\ref{eq:problem}). The Monte Carlo estimator of problem (\ref{eq:problem}), denoted by $\Eh[\cdot]$, is given by
\begin{align}
\Ee[\M]=\E \left[\dfrac{1}{T+1}\sum_{t=0}^T \| z(t)-\zh(t) \|^2 \right]\approx\nonumber\\ \dfrac{1}{K(T+1)}\sum_{k=1}^K\sum_{t=0}^T \|z^k(t)-\zh^k(t)\|^2\ \text{\reflectbox{$\coloneqq$}}\ \Eh[\M] .\label{eq:Eh_def}
\end{align}

\begin{figure}[t]
\centering
\begin{tikzpicture}
\draw (0,4.75) node[above]{$\M$};
\draw [->,>=latex] (0,4.75) -- (0,4.25);
\draw (-1.5,3.75) rectangle (1.5,4.25);
\draw (0,4) node{Draw};
\draw [->,>=latex] (-1,3.75) -- (-1,2.25);
\draw (-1,3) node[left]{$\{x^k(t)\}^{k=1,\dots,K}_{t=0,\dots,T}$};
\draw [->,>=latex] (1,3.75) -- (1,2.25);
\draw (1,3) node[right]{$\{v^k(t)\}^{k=1,\dots,K}_{t\in\M}$};
\draw (0,2) node{Model};
\draw (-1.5,1.75) rectangle (1.5,2.25);
\draw [->,>=latex] (-1,1.75) -- (-1,0.25);
\draw (-1,0) node{PF};
\draw (-1.5,-0.25) rectangle (-0.5,0.25);
\draw (-1,1) node[left]{$\{y^k(t)\}^{k=1,\dots,K}_{t\in\M}$};
\draw [->,>=latex] (-1,-0.25) -- (-1,-1.75);
\draw (-1.5,-2.25) rectangle (1.5,-1.75);
\draw (-1,-1) node[left]{$\{\zh^k(t)\}^{k=1,\dots,K}_{t=0,\dots,T}$};
\draw (0,-2) node{MSE};
\draw [->,>=latex] (1,1.75) -- (1,-1.75);
\draw (1,0) node[right]{$\{z^k(t)\}^{k=1,\dots,K}_{t=0,\dots,T}$};
\draw [->,>=latex] (0,-2.25) -- (0,-2.75);
\draw (0,-2.75) node[below]{$\Eh[\M]$};
\end{tikzpicture}
\caption{Representation of the Monte Carlo algorithm which computes $\M\mapsto \Eh[\M]$. The outputs generated by the Draw block are drawn according to (\ref{eq:model:x}), (\ref{eq:model:x0}) and the distribution of $v(t)$. The model block uses respectively equations (\ref{eq:model:y}) and (\ref{eq:model:z}) to compute the $y^k(t)$ and $z^k(t)$ that correspond to the $x^k(t)$ and the $v^k(t)$ received in inputs. The PF block represents a particle filter that computes an estimate $\zh^k(t)$ of $z^k(t)$ from previous intermittent measurements $\{y^k(\tau)\}_{\tau\leq t,\tau\in\M}$. The MSE block computes the mean squared of the difference between the inputs, it is $\Eh[\M]$ defined in equation (\ref{eq:Eh_def}).}
\label{fig:schemeMC}
\end{figure}
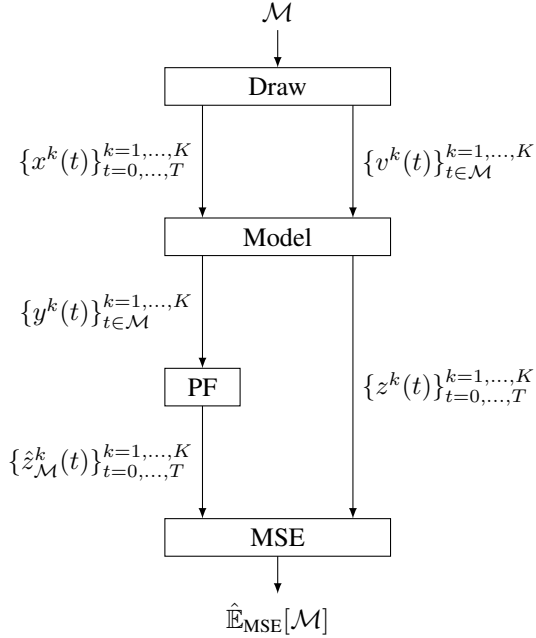

With this notation, problem (\ref{eq:problem}) is approximately equivalent to
\begin{align}\label{eq:problem:approx}
\min_{\M\subseteq\{0,\dots,T\}} \Eh[\M ] \text{\ subject\ to\ } |\M |=N.
\end{align}
The fact that we only have access to an approximation of the objective function requires a particular attention during the optimization. The nesting of the Monte Carlo algorithm and the particle filter is represented in figure \ref{fig:schemeMC}.

\begin{figure}[t]
    \centering
\includegraphics[width=8.5cm]{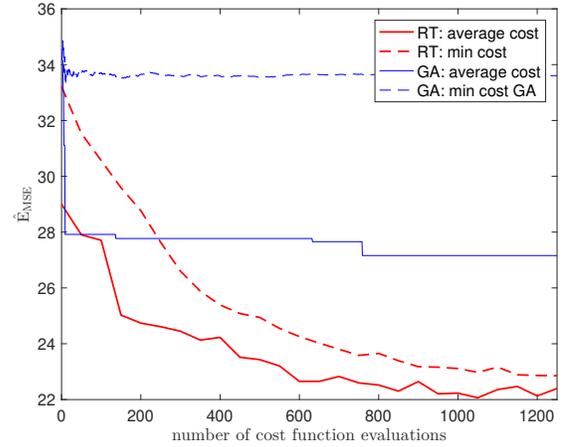}
    \caption{Evolution of the average and minimum cost $\Eh$ with respect to the number of cost function evaluations for both the genetic algorithm (GA) and the random trials optimizer (RT). It illustrates the quality of minimization for given computational resources. One generation of the genetic algorithm corresponds to 50 evaluations.}
    \label{Genetic algo :min and average cost}
\end{figure}

\subsection{Genetic algorithm}\label{sec:ga}
Solving the combinatorial optimization problem (\ref{eq:problem:approx}) corresponds to finding the set $\M\subseteq\{0,\dots,T\}$ of cardinality $N$ that minimizes $\Eh[\M]$. An exhaustive search would require to test all admissible $\M$ which represents $\frac{(T+1)!}{(T+1-N)!N!}$ possibilities. It is computationally intractable for large $N$ and $T$.

A genetic algorithm \cite{mitchell1998introduction} is used to find an approximate solution of problem (\ref{eq:problem:approx}). Genetic algorithms are generally used for unconstrained optimization. In our case, to deal with constraint $|\M|=N$, a count preserving crossover is implemented \cite{umbarkar2015crossover}.

If measurement times are widely spaced for an individual of the genetic algorithm, it can happen that all particles of the particle filter have zero weights (due to degeneracy problem \cite{arulampalam2002tutorial}). In such a case, the corresponding individual is killed and will not be used for next generations.

As mentioned in the previous section, the objective function $\Ee[\M]$ can only be evaluated approximately as $\Eh[\M]$. It makes the minimization difficult if too sensitive to bad cost function estimations. To face this issue, our genetic algorithm returns the best individual of the last generation instead of the best individual among all generations.

In Section \ref{sec:resultsAndDiscussions}, the performance of the genetic algorithm is compared to a random trial optimizer. It will show that the ability of evolutionary algorithms to broadly sample a population makes the genetic algorithm better suited for the addressed problem.

Our implementation of the genetic algorithm uses stochastic universal sampling and sigma scaling with unitary sigma coefficient \cite{mitchell1998introduction}. Crossover probability is 1 and mutation probability per gene is 0.003.



\section{Results and discussion}\label{sec:resultsAndDiscussions}
In order to illustrate the performances obtained by our approach, the following commonly studied model is used \cite{arulampalam2002tutorial,carlin1992monte,kitagawa1996monte,kadirkamanathan2002particle}, 
\begin{alignat}{3}
x(t+1) &=&&\ \dfrac{x(t)}{2} + \frac{25 x(t)}{1+x(t)^2} + 8 \cos(1.2 t) + w(t),  \label{eq:toy_ex:x}\\
y(t)   &=&&\ \frac{x(t)^2}{20} + v(t) \text{\ \ for\ \ }  t\in\M, \label{eq:toy_ex:y}\\
z(t)   &=&&\ x(t), \label{eq:toy_ex:z}\\
x(0) &\sim&&\ \N(0,5^2), \label{eq:toy_ex}
\end{alignat}
where $w(t)\sim\N(0,1)$ and $v(t)\sim\N\left(0,(\sin(0.25t) + 2)^2\right)$ are zero mean independent Gaussian noise processes. Equations (\ref{eq:toy_ex:x}) and (\ref{eq:toy_ex:z}) hold for $t=0,\dots,T-1$ and $t=0,\dots,T$, respectively.

The problem is to find the best set $\M$ of $N=21$ measurement times in the range of 0 to $T=60$ such that $\Eh[\M]$ is minimized. Even for a problem this size, an exhaustive search would require to test over $1.2 \cdot 10^{15}$ admissible $\M$.

In the following, the particle filter uses 500 particles, the Monte Carlo algorithm uses $K=1,000$ draws, and the population size and the number of generations of the genetic algorithm are respectively 50 and 25.

Figure \ref{Genetic algo :min and average cost} illustrates the evolution of the average and minimum $\Eh$ with respect to the number of cost function evaluations (one generation of the genetic algorithm corresponds to 50 evaluations, i.e. the population size). The average and minimum $\Eh$ decrease over generations until reaching quasi-convergence. As the genetic algorithm has reached a quasi-convergence state, most of the individuals are identical, which shows good convergence behaviour. The set of measurement times returned by the genetic algorithm is denoted $\M_{\text{GA}}$.

In addition, the genetic algorithm is compared to a random trial optimization method. It samples measurement times randomly and evaluates their corresponding costs. Figure \ref{Genetic algo :min and average cost} indicates the average and minimum costs of the random trials with respect to the number of cost function evaluations, i.e. the number of trials. One can observe that for a same computational cost, i.e. a same number of cost function evaluations, our genetic algorithm significantly outperforms the random trial optimizer.



\begin{figure}[t]
    \centering
    \includegraphics[width=8.5cm]{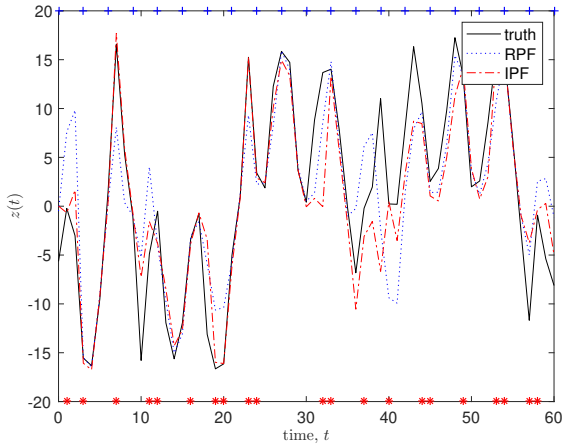}
    \caption{Comparison of true value $z(t)$ with the values $\zh(t)$ filtered by the regular particle filter (RPF) and the optimal intermittent particle filter (IPF) over a single draw. The relative gain on the complete sequence is $g=28.3\%$. Optimal measurement times $\M_{\text{GA}}$ are indicated with red stars (*) and regularly spaced measurement times $\M_{\text{REG}}$ are indicated with blue plus (+). Results simulated from model (\ref{eq:toy_ex:x})-(\ref{eq:toy_ex}).}
    \label{fig:Toy_ex:draw}
\end{figure}

Now that the genetic algorithm performance has been demonstrated, our \textit{optimal intermittent particle filter} (IPF) method is compared with a \textit{regular particle filter} (RPF). It is a particle filter with regularly spaced measurement times, defined by
\begin{align}\label{eq:M_reg}
\M_\text{REG}\coloneqq\left\{\left. \text{Round}\left[\dfrac{kT}{N-1}\right] \right| k=0,\dots,N-1  \right\},
\end{align}
where $\text{Round}[\cdot]$ is the rounding operator.

For a given measurement times set $\M$, one defines the random variable $\text{MSE}[\M] \coloneqq \frac{1}{T+1} \sum_{t=0}^{T} \| z(t) - \hat{z}_{\M}(t) \|^2$. With this notation, one can define the relative gain, $g \coloneqq \frac{\text{MSE}[\M_{\text{REG}}]- \text{MSE}[\M_{\text{GA}}]}{\text{MSE}[\M_{\text{REG}}]}$. This relative gain is positive when our IPF method outperforms the RPF.

Figure \ref{fig:Toy_ex:draw} compares the estimates produced by RPF and IPF to the exact $z(t)$ for one particular realization. Optimal measurement times $\M_{\text{GA}}$ and regularly spaced measurement times $\M_{\text{REG}}$ are indicated with red stars (*) and blue plus (+), respectively. One can observe better filtering performances using IPF. Quantitatively, the relative gain obtained using IPF instead of RPF is $g=28.3\%$. Due to the nonlinearities of the problem, the obtained measurement times can not be easily interpreted.

\begin{figure}[t]
    \centering
    \includegraphics[width=8.5cm]{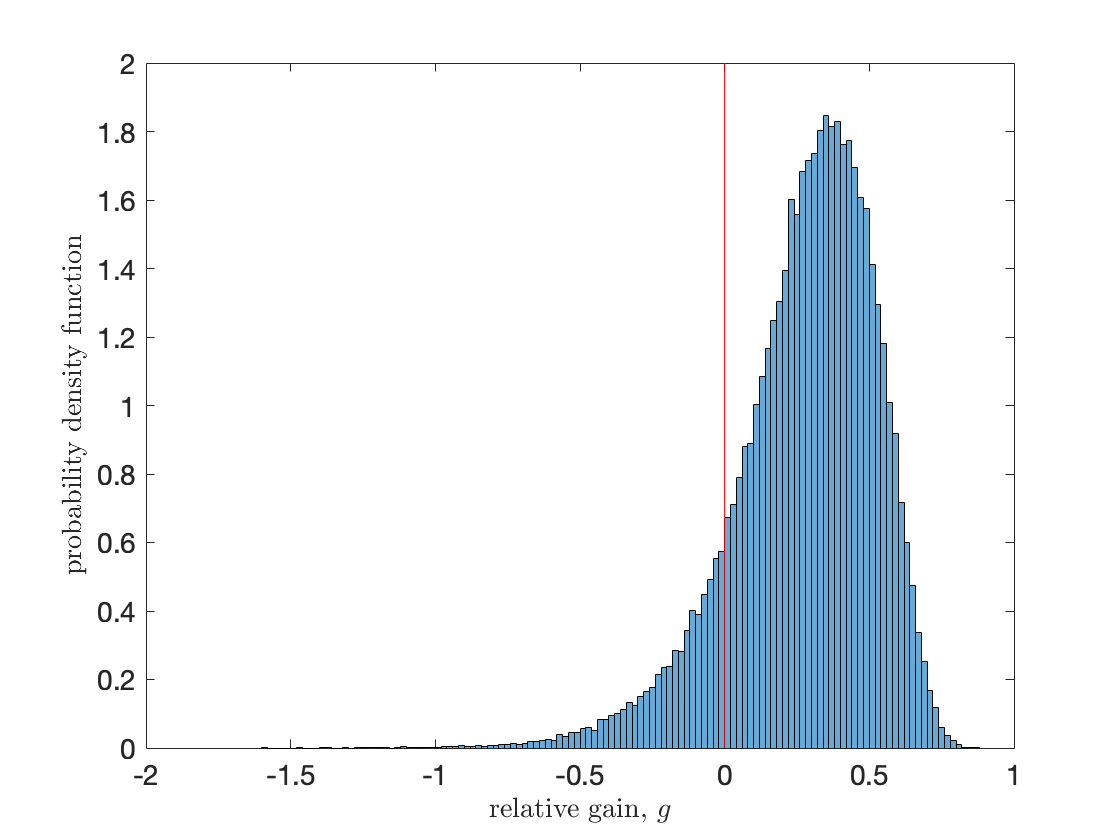}
    \caption{Histogram of the relative gain $g = \frac{\text{MSE}[\M_{\text{REG}}]- \text{MSE}[\M_{\text{GA}}]}{\text{MSE}[\M_{\text{REG}}]}$. It has been obtained by running both the regular particle filter and the intermittent particle filter on the same $100,000$ draws. The vertical red line corresponds to a null gain. The average is $27.7\%$ and $g$ is positive in $87.5\%$ of the cases.}
    \label{fig:PDF_gain}
\end{figure}

The histogram presented in figure \ref{fig:PDF_gain} approximates the probability density function of the relative gain $g$. It is obtained by running on 100,000 draws both the IPF and the RPF and computing the corresponding gain $g$. The mean relative gain is $27.7\%$ and in $87.5\%$ of the cases, our IPF method outperformed the RPF.



\section{Conclusion}\label{sec:conclusion}
The problem of selecting optimal measurement times for particle filtering over a finite time horizon was presented. Then, an algorithm nesting a genetic algorithm, a Monte Carlo method and a particle filter was proposed to find these optimal measurement times.

Firstly, a numerical example demonstrated that our genetic algorithm significantly outperforms a random trial optimiser. Then, we demonstrated that in comparison to regularly spaced measurements, the optimal choice of intermittent measurement times led to better filtering performance in $87.5\%$ of the cases. On average, the relative gain was $27.7\%$.

Further work will consider selecting the measurement times \textit{online} instead of fixing them \textit{a priori}: after measurements have been acquired, one can recompute the next optimal measurement times, thereby incorporating all the then-available information. Extensions to continuous-time systems with discrete measurements will also be investigated. Finally, robustness analyses will be performed to determine how model uncertainties affect the performance of the proposed method, for situations in which the dynamics of the system are only approximately known.




Overall, our results demonstrate the added value of \textit{a priori} selecting measurement times based on system dynamics for the optimal estimation of a state from limited measurements.


\bibliographystyle{IEEEbib}
\bibliography{refs}
\end{document}